\documentclass[12pt]{article}
\usepackage{epsfig}
\newcommand{\bea}{\begin{eqnarray}}
\newcommand{\eea}{\end{eqnarray}}
\newcommand{\be}{\begin{equation}}
\newcommand{\ee}{\end{equation}}

\title{ Application of the rescaling model at small Bjorken $x$ values
  \author{  A.V.~Kotikov$^{1,3}$, B.G.~Shaikhatdenov$^{3}$,Pengming Zhang$^{1,2}$ \\
    $^{1}$ Institute of Modern Physics, Lanzhou 730000, China\\
    $^{2}$ University of Chinese Academy of Sciences, Yuquanlu 19A, Beijing 100049, China\\
 $^{3}$ Joint Institute for Nuclear Research, 141980, Dubna, Russia } }
 
\begin{document}
\maketitle
\abstract{
The Bessel-inspired behavior of parton densities at small Bjorken $x$ values,
obtained in the case of the flat initial conditions for DGLAP evolution equations,
is used along with ``frozen'' and analytic modifications of the
strong coupling constant to study the so-called EMC effect.
Among other results, this approach allowed predicting small $x$ behavior of the gluon density in nuclei.}\\

{\it Keywords:} Deep inelastic scattering; parton densities; EMC effect.

\section{Introduction}
The study of deep-inelastic scattering (DIS) of leptons off nuclei reveals
an appearance of a significant nuclear effect (for a review see, e.g.,~\cite{Arneodo:1992wf,Rith:2014tma}).
It was first observed by the European Muon Collaboration~\cite{Aubert:1983xm} in the valence quark dominance region;
hence the name. This observation rules out the naive picture of a nucleus as being a system of quasi-free nucleons.

There in general are two mainstream approaches to studying the EMC effect. In the first one, which is at present 
more popular, nuclear parton distribution functions (nPDFs) are extracted from the global fits to nuclear
data by using empirical parametrizations of their normalizations (see~\cite{Eskola:2009uj,Eskola:2016oht,Khanpour:2016pph}).
This is completely analogous to respective studies of usual (nucleon) PDFs (see recent analyses in~\cite{fits}).
Both PDFs and nPDFs are obtained from the numerical solution to Dokshitzer-Gribov-Lipatov-Altarelli-Parisi (DGLAP)
equations~\cite{DGLAP}
\footnote{Sometimes, in the analyses of DIS experimental data it is convenient to use an exact
  solution to DGLAP equations in the Mellin moment space and reconstruct SF $F_2$ from the moments
 (see recent paper~\cite{Kotikov:2016ljf} and references and discussions therein). The studies of
  nuclear effects in such a type of analyses can be found in~\cite{Krivokhizhin:2005pt}, though its
  consideration is beyond the scope of the present study.}.
The second strategy is based upon some models of nuclear PDFs (see different models in,
for example,~\cite{Kulagin:2004ie}--\cite{Close:1984zn}
and a recent review~\cite{Kulagin:2016fzf}).

Here we will follow the rescaling model~\cite{Jaffe:1983zw,Close:1984zn}, which was very popular some time ago.
The model is based on a suggestion~\cite{Close:1983tn} that the effective confinement size of gluons and quarks in the
nucleus is greater than in a free nucleon. In the framework of perturbative QCD it was found \cite{Jaffe:1983zw,Close:1984zn,Close:1983tn}
that such a change in the confinement scale predicts that nPDFs and PDFs can be related by simply rescaling their arguments
(see Eq.~(\ref{va.1a}) below). Thus, in a sense, the rescaling model lies in-between two above approaches: in its framework there 
are certain relations between usual and nuclear PDFs that result from shifting the values of kinematical variable $\mu^2$;
however, both densities obey DGLAP equations.

At that time, the model was established for the valence quark dominance region $0.2 \leq x \leq 0.8$.
The aim of our paper is to extend its applicability to the region of small $x$ values,
where the rescaling values can be different for gluons and quarks. To see it clearly
we use the generalized double-scaling approach (DAS)~\cite{Munich,Q2evo}. The latter is
based upon the analytical solution to DGLAP equations in the small $x$ region and generalizes
earlier studies~\cite{BF1}.

A few years ago most analyses of nPDFs have been done in the leading order
(LO) of perturbation theory, but now the situation is drastically changed and the standard level of accuracy in current
analyses is at the next-to-leading order (NLO) one (see~\cite{Eskola:2009uj,Eskola:2016oht}).
Even more, there have already appeared a global analysis~\cite{Khanpour:2016pph} 
performed at the next-to-next-to-leading order.
Nevertheless the present analysis will be carried out in LO. We note that the analysis to this
level of accuracy is just for the start and can be considered as a first step in our investigations
in this direction. We are going to improve the accuracy at least to the NLO level in the future works.

\section{SF $F_2$ at low $x$}

A reasonable agreement between HERA data~\cite{H1ZEUS} and predictions made
by perturbative Quantum Chromodynamics (QCD) was observed for $Q^2 \geq 2$ GeV$^2$ 
\cite{CoDeRo}, thereby promising that perturbative QCD is capable of describing the
evolution of parton densities down to very low $Q^2$ values.

Some time ago ZEUS and H1 Collaborations have presented new precise combined data~\cite{Aaron:2009aa}
on the structure function (SF) $F_2$.
An application of the generalized DAS approach~\cite{Q2evo} at NLO
%the next-to-leading order (NLO)
shows that theoretical predictions are well compatible with experimental data at $Q^2 \geq 3\div 4$ GeV$^2$
(see recent results in~\cite{Kotikov:2012sm}).

In the present paper we perform a LO analysis of the combined data~\cite{Aaron:2009aa}
where the SF $F_2$ has the following form
\begin{eqnarray}
  F_2(x,\mu^2) &=& e \,
  %\bigl(
  f_q(x,\mu^2),
  %+ \frac{2}{3}f a_s(Q^2)f_g(x,Q^2)\bigr),
%  ~~ a_s(\mu^2) = \frac{\alpha_s(\mu^2)}{4\pi},  
%\nonumber \\
\label{8a} 
 \end{eqnarray}
where
$e=(\sum_1^f e_i^2)/f$ is an average of the squared quark charges.
Notice that the approach used in these analyses will be analogous to that exploited in NLO ones carried out 
in~\cite{Kotikov:2012sm}--\cite{Cvetic1}.

The small-$x$ asymptotic expressions for parton densities $f_a$ can be written as follows
\begin{eqnarray}
f_a(x,\mu^2) &=& 
%~=~ 
f_a^{+}(x,\mu^2) + f_a^{-}(x,\mu^2),~~(\mbox{hereafter}~~~a=q,g) \nonumber \\
	f^{+}_g(x,\mu^2) &=& \biggl(A_g + \frac{4}{9} A_q \biggl)
		\tilde{I}_0(\sigma) \; e^{-\overline d_{+} s} + O(\rho), \nonumber \\
f^{+}_q(x,\mu^2) &=& 
%&=& 
\frac{f}{9} \biggl(A_g + \frac{4}{9} A_q \biggl) \rho \tilde{I}_1(\sigma)  \; e^{-\overline d_{+} s} 
+ O(\rho),
	\label{8.01} \\
%\nonumber \\
        f^{-}_g(x,\mu^2) &=& -\frac{4}{9} A_q e^{- d_{-} s} \, + \, O(x),~~
        %\nonumber \\
%	\label{8.00} \\
	f^{-}_q(x,\mu^2) ~=~  A_q e^{-d_{-}(1) s} \, + \, O(x),
	\label{8.02}
\end{eqnarray}
where $I_{\nu}$ ($\nu=0,1$)
%$\tilde{I}_{\nu}$ ($\nu=0,1$)
are the modified Bessel functions
%(at $s\geq 0$,
%i.e. $\mu^2 \geq\mu^2_0$) and the usual Bessel functions (at $s< 0$)
%\begin{equation}
%\rho^{\nu} \tilde{I}_{\nu}(\sigma) =
%\left\{
%\begin{array}{ll}
%\rho^{\nu} I_{\nu}(\sigma) , & \mbox{ if } s \geq 0; \\
%(-\tilde{\rho})^{\nu} J_{\nu}(\tilde{\sigma}) , & \mbox{ if } s < 0. 
%\end{array}
%\right. \, ,~~
%\rho^{-\nu} \tilde{I}_{\nu}(\sigma) =
%\left\{
%\begin{array}{ll}
%\rho^{-\nu} I_{\nu}(\sigma) , & \mbox{ if } s \geq 0; \\
%\tilde{\rho}^{-\nu} J_{\nu}(\tilde{\sigma}) , & \mbox{ if } s < 0. 
%\end{array}
%\right.
%\label{4}
%\end{equation}
with
\bea
%\begin{equation}
&&s=\ln \left( \frac{a_s(\mu^2_0)}{a_s(\mu^2)} \right),~~
\sigma = 2\sqrt{\left|\hat{d}_+\right| s
  \ln \left( \frac{1}{x} \right)}  \ ,~~~ \rho=\frac{\sigma}{2\ln(1/x)},~~ \nonumber \\
&&a_s(\mu^2) \equiv \frac{\alpha_s(\mu^2)}{4\pi} = \frac{1}{\beta_0\ln(\mu^2/\Lambda^2_{\rm LO})}
%,~~
%\tilde{\sigma} = 2\sqrt{-\left|\hat{d}_+\right| s
%  \ln \left( \frac{1}{x} \right)}  \ ,~~~ \tilde{\rho}=\frac{\tilde{\sigma}}{2\ln(1/x)}
\label{intro:1a}
\eea
%\end{equation}
and 
\begin{equation}
\hat{d}_+ = - \frac{12}{\beta_0},~~~
\overline d_{+} = 1 + \frac{20f}{27\beta_0},~~~ 
d_{-} = \frac{16f}{27\beta_0}
\label{intro:1b}
\end{equation}
denote singular $\hat{d}_+$ and regular $\overline d_{+}$ parts of the ``anomalous dimensions'' 
$d_{+}(n)$ and $d_{-}(n)$
\footnote{Note that the variables $d_{\pm}(n)$ are ratios $\gamma_{\pm}^{(\rm LO)}(n)/(2\beta_0)$
  of LO anomalous dimensions $\gamma_{\pm}^{(\rm LO)}(n)$ and LO coefficient $\beta_0$ of QCD
  $\beta$-function.},
respectively, in the limit $n\to1$.

By using the expressions given above we have analyzed H1 and ZEUS data for $F_2$ \cite{Aaron:2009aa}.
In order to keep the analysis as simple as possible, here we take $\mu^2=Q^2$ and $\alpha_s(M^2_Z)=0.1168$ 
in agreement with ZEUS results presented in~\cite{H1ZEUS}.
Moreover, we use the fixed flavor scheme with two different values $f=3$ and  $f=4$ of active quarks.

As can be seen from Table 1, the twist-two approximation looks reasonable for $Q^2 \geq 3.5$ GeV$^2$.
It is almost completely compatible with NLO analyses done in~\cite{Kotikov:2012sm}--\cite{Cvetic1}.
Moreover, these results are rather close to original analyses
(see~\cite{Cooper-Sarkar:2016foi} and references therein) performed by the HERAPDF group.
As in the case of~\cite{Cooper-Sarkar:2016foi} our $\chi^2/DOF \sim 1$ unless combined H1 and ZEUS experimental 
data analyzed are kept according to $Q^2 \geq 3.5$ GeV$^2$.

At lower $Q^2$ there is certain disagreement, which is we believe to be explained by the higher-twist (HT) 
corrections playing their important role. These HT corrections have rather cumbersome form
at low $x$~\cite{HT}. As it was shown~\cite{Cvetic1}, it is very promising to use
infrared modifications of the strong coupling constant in our analysis.
Such types of coupling constants modify the low $\mu^2$ behavior of parton densities
and structure functions. What is important, they do not generate additional free parameters.
Moreover, the present results will be applied in the analyses of NMC data (see Sect.~5 and~6) accumulated
at very low $Q^2$ values, where the HT expansion ($\sim 1/Q^{2n}$) is thought to be not applicable.

So, following~\cite{Cvetic1}, we are going to use the so-called ``frozen'' $a_{\rm fr}(\mu^2)$ \cite{Badelek:1996ap}
and analytic $a_{\rm an}(\mu^2)$ \cite{Shirkov:1997wi} versions
\be
a_{\rm fr}(\mu^2) = a_s(\mu^2 + M^2_{g}),~~~
%\label{Fro} \\&&
a_{\rm an}(\mu^2) = a_s(\mu^2) - \frac{1}{\beta_0} \, \frac{\Lambda_{\rm LO}^2}{\mu^2-\Lambda_{\rm LO}^2} \, ,
\label{Ana}
\ee
where $M_{g}$ is a gluon mass with $M_{g}$=1 GeV$^2$ (see~\cite{Shirkov:2012ux} and references therein
\footnote{There are a number of various approaches to define the value of this gluon mass and even
the form of its momentum dependence (see, e.g., a recent review~\cite{Deur:2016tte}).}).

It is seen that the results of the fits carried out when $a_{\rm fr}(\mu^2)$ and $a_{\rm an}(\mu^2)$ are used,
are very similar to the corresponding ones obtained in~\cite{Kotikov:2012sm}. 
Moreover, note that the fits in the cases with ``frozen'' and analytic strong coupling constants look 
very much alike (see also~\cite{Cvetic1,KoLiZo}) and describe fairly well the data in the low $Q^2$ region,
as opposed to the fits with a standard coupling constant, which largely fails here.
The results are presented in Table~1. With the number of active quarks $f=4$, they are shown also in Fig.~1.

Just like the previous analyses~\cite{Kotikov:2012sm,Cvetic1,KoLiZo} we observe strong improvement 
in the agreement between theoretical predictions and experimental data once ``frozen'' and analytic 
modifiations to the coupling constant are applied. When the data are cut by $Q^2 \geq $ 1 GeV$^2$,
$\chi^2$ value drops by more than two times. Ditto for the analyses of data with $Q^2 \geq $ 3.5 GeV$^2$ imposed.\\

{\bf Table 1.}
\begin{center}
\begin{tabular}{|c|c|c|c|c|c|c|}
\hline
 $f=3$        &  $a_s(Q^2)$ &  $a_s(Q^2)$ & $a_{an}(Q^2)$ & $a_{an}(Q^2)$ & $a_{fr}(Q^2)$ &  $a_{fr}(Q^2)$       \\
$Q^2 \geq $ & 1 GeV$^2$ & $3.5$ GeV$^2$ & $1$ GeV$^2$ & $3.5$ GeV$^2$ & $1$ GeV$^2$ & $3.5$ GeV$^2$   \\
%&               &      &                           &    &          \\
\hline
%\hline
$A_g$       &  0.46 $\pm$ 0.02  & 0.74 $\pm$ 0.04 & 1.16 $\pm$ 0.03 &  1.30 $\pm$ 0.04 & 0.96 $\pm$ 0.03 &
1.06 $\pm$ 0.04  \\
$A_q$       &  1.58 $\pm$ 0.04  & 1.48 $\pm$ 0.06 & 1.16 $\pm$ 0.04 &  1.21 $\pm$ 0.07 & 1.23 $\pm$ 0.08 &
1.32 $\pm$ 0.07  \\
$Q_0^2$       &  0.40 $\pm$ 0.01  & 0.46 $\pm$ 0.01 & 0.20 $\pm$ 0.01 &  0.16 $\pm$ 0.01 & 0.49 $\pm$ 0.01 &
0.53 $\pm$ 0.01  \\
$\chi^{2}$ & 365.7 &  69.7 & 149.7  & 42.9 & 140.4 & 47.6 \\
\hline \hline
 $f=4$        &  $a_s(Q^2)$ &  $a_s(Q^2)$ & $a_{an}(Q^2)$ & $a_{an}(Q^2)$ & $a_{fr}(Q^2)$ &  $a_{fr}(Q^2)$       \\
$Q^2 \geq $ & 1 GeV$^2$ & $3.5$ GeV$^2$ & $1$ GeV$^2$ & $3.5$ GeV$^2$ & $1$ GeV$^2$ & $3.5$ GeV$^2$   \\
%&               &      &                           &    &          \\
\hline
%\hline
$A_g$       &  0.47 $\pm$ 0.02  & 0.54 $\pm$ 0.03 & 0.65 $\pm$ 0.02 &  0.76 $\pm$ 0.03 & 0.96 $\pm$ 0.03 &
0.77 $\pm$ 0.03  \\
$A_q$       &  1.58 $\pm$ 0.04  & 1.09 $\pm$ 0.06 & 0.95 $\pm$ 0.03 &  0.96 $\pm$ 0.04 & 1.23 $\pm$ 0.05 &
0.95 $\pm$ 0.06  \\
$Q_0^2$       &  0.40 $\pm$ 0.01  & 0.37 $\pm$ 0.01 & 0.16 $\pm$ 0.01 &  0.19 $\pm$ 0.01 & 0.49 $\pm$ 0.01 &
0.43 $\pm$ 0.01  \\
$\chi^{2}$ & 366.0 &  57.0 & 166.3  & 43.6 & 140.0 & 40.6 \\
\hline 
\end{tabular}
\end{center}

\begin{figure}[t]
\centering
\vskip 0.5cm
\includegraphics[height=0.45\textheight,width=0.8\hsize]{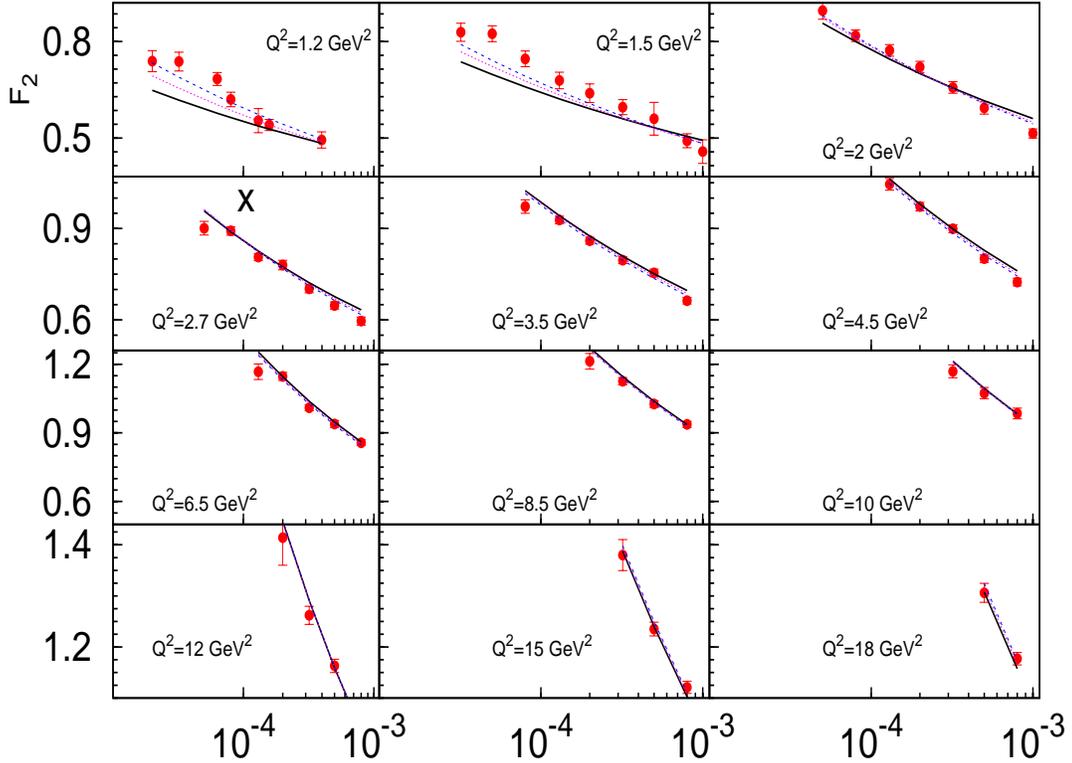}
\vskip -0.3cm
\caption{$x$ dependence of $F_2(x,Q^2)$ in bins of $Q^2$.
The combined experimental data from H1 and ZEUS Collaborations 
\cite{Aaron:2009aa} are
compared with the LO fits for $Q^2\geq1$~GeV$^2$ implemented with a standard strong coupling constant (solid lines),
and its frozen (dash-dotted lines) and analytic (dashed lines) modifications.}
\label{fig:F1}
\end{figure}

Recent NLO analyses (see the third paper in~\cite{Kotikov:2012sm})
have been carried out within the framework of the fixed flavor scheme with $f=3$ active
light flavors and with a purely perturbative charm quark generated in a photon-gluon fusion (PGF) process.
Such type of analyses for the complete SF $F_2(x,Q^2)$ cannot be done at LO. 
\footnote{Notice that the SF $F_{2c}(x,Q^2)$, the charm part of $F_2(x,Q^2)$, appears with $a_s(Q^2)$ and can be confronted
already at LO with the data produced in a PGF process (see Sect. 7 below).}

Therefore, we should use some fixed values of active quarks.
Nevertheless, we would like to note that the results obtained here and those in~\cite{Kotikov:2012sm}---\cite{Cvetic1},
where various schemes were used, are very stable and close to each other.

\section{Rescaling model}

In the rescaling model~\cite{Close:1984zn} SF $F_2$ and, therefore, valence part of quark densities,
gets modified in the case of a nucleus $A$ at intermediate and large $x$ values $(0.2 \leq x \leq 0.9)$ as follows
\begin{equation}
%\begin{eqnarray}
  F_2^A(x,\mu^2) =
  %&=&
  F_2(x,\mu^2_{A,v}),~~~
  %\nonumber \\
  f_{NS}^A(x,\mu^2) =
  %&=&
  f_{NS}(x,\mu^2_{A,v}),
  \label{va.1}
%\end{eqnarray}
\end{equation}
where a new scale $\mu^2_{A,v}$ is related with $\mu^2$ as
\begin{equation}
  \mu^2_{A,v} = \xi^A_v(\mu^2)\mu^2,~~~
   \xi^A_v(\mu^2) = {\left(\frac{\lambda_A^2}{\lambda_N^2}\right)}^{a_s(\tilde{\mu}^2)/a_s(\mu^2)} \, 
  \label{va.1a}
\end{equation}
where some additional scale $\tilde{\mu}^2=0.66$ GeV$^2$, which was in its turn an initial point
in a $\mu^2$-evolution performed in~\cite{Close:1984zn}; it is then estimated in Appendix~A of that paper.
The quantity $\lambda_A/\lambda_N$ stands for the ratio of quark confinement radii in a nucleus $A$ and nucleon.
The values of $\lambda_A/\lambda_N$ and $\xi^A_v(\mu^2)$ at $\mu^2=20$ GeV$^2$
were evaluated for different nuclei and presented in Tables I and II in~\cite{Close:1984zn}.

Since the factor $ \xi^A_v(\mu^2)$ is $\mu^2$ dependent, it is convenient to transform it
to some $\mu^2$ independent one. To this end, we consider the variable
$\ln(\mu^2_{A,v}/\Lambda^2)$, which has the following form (from Eq.~(\ref{va.1a}))
\begin{equation}
\ln\left(\frac{\mu^2_{A,v}}{\Lambda^2}\right) = \ln\left(\frac{\mu^2}{\Lambda^2}\right) \cdot \Bigl(1+ \delta^A_v\Bigr)
%  \xi^A_v(Q^2) = {\left(\frac{\lambda_N}{\lambda_A}\right)}^{a_s(\mu^2)/a_s(\mu_A^2)} \,,
  \label{va.1b}
\end{equation} 
where the nuclear correction factor $\delta^A_v$  becomes $\mu^2$ independent:
\begin{equation}
  \delta^A_v = \frac{1}{\ln\left(\tilde{\mu}^2/\Lambda^2\right)} \,
  %\frac{1}{\ln\left(\frac{\tilde{\mu}^2}{\Lambda^2}\right)} \,
  \ln\left(\frac{\lambda_A^2}{\lambda_N^2}\right)\,,
\label{delta}
\end{equation}
where it is seen that two parameters, namely, the scale $\tilde{\mu}$ and ratio $\lambda_A/\lambda_N$,
are combined to form a $Q^2$-independent quantity.
Using Eqs.~(\ref{va.1b}) and/or~(\ref{delta}), we can recover results for $\delta^A_v$, which are presented in Table~2.

{\bf Table 2.}
\begin{center}
\begin{tabular}{|c|c|c|c|c|c|c|}
\hline
%        &               &    &  &                       &        &         \\
$A$ & ${}^2$D & ${}^4$He & ${}^7$Li & ${}^{12}$C &  ${}^{40}$Ca    \\
$N$ &             & 11   & 16        & 16              & 11      \\
%&               &      &                           &    &          \\
\hline
%\hline
$\delta^A_v$       &    0.01 &  0.06   & 0.05  & 0.08 & 0.11  \\
$\delta^{AD}_v$       &    0 &  0.05   & 0.04  & 0.07 & 0.10  \\
%%-$\delta^{AD}_+$   &  0  & 0.07 $\pm$ 0.01 & 0.05 $\pm$ 0.01 &  0.11 $\pm$ 0.01  & 0.21 $\pm$ 0.01 & 0.22 $\pm$ 0.11 \\
-$\delta^{AD}_{+,an}$   &  0  & 0.06 $\pm$ 0.01 & 0.06 $\pm$ 0.01 &  0.11 $\pm$ 0.01  & 0.19 $\pm$ 0.01   \\
-$\delta^{AD}_{-,an}$   &  0  & 0.24 $\pm$ 0.08 & 0.22 $\pm$ 0.07 &  0.41 $\pm$ 0.04  & 0.51 $\pm$ 0.04   \\
$\chi^{2}_{an}$   &  0       &    4.68    & 17              &  9.68                   & 12    \\
-$\delta^{AD}_{+,fr}$   &  0  & 0.06 $\pm$ 0.01 & 0.06 $\pm$ 0.01 &  0.12 $\pm$ 0.01  & 0.21 $\pm$ 0.02   \\
-$\delta^{AD}_{-,fr}$   &  0  & 0.32 $\pm$ 0.08 & 0.28 $\pm$ 0.07 &  0.54 $\pm$ 0.04 & 0.71 $\pm$ 0.04   \\
$\chi^{2}_{fr}$   &  0  &    5      & 35                &  26            & 37   \\
\hline 
\end{tabular}
\end{center}

Since our parton densities contain the variable $s$ defined in Eq.~(\ref{intro:1a}),
it is convenient to consider its $A$ modification. It has the following simple form:
\begin{equation}
s^A_v \equiv \ln \left(\frac{\ln\left(\mu^2_{A,v}/\Lambda^2\right)}{\ln\left(\mu^2_{0}/\Lambda^2\right)}\right)
= s +\ln\Bigl(1+\delta^A_v\Bigr) \approx s +\delta^A_v,~~~ 
\label{sA}
\end{equation}
i.e. the nuclear modification of the basic variable $s$ depends on the 
$\mu^2$ independent parameter $\delta^A_v$, which possesses very small values.

\section{Rescaling model al low $x$}

Standard evidence coming from earlier studies contains conclusion about inapplicability
of the rescaling model at small $x$ values (see, for example,~\cite{Efremov:1986mt}).
It looks like it can be related with some simplifications of low $x$ analyses (see, for example,~\cite{Kotikov:1988aa},
where the rise in EMC ratio was wrongly predicted at small $x$ values).

Using an accurate study of DGLAP equations at low $x$ within the framework of the generalized DAS approach, 
it is possible to achieve nice agreement with
the experimental data for the DIS structure functon $F_2$ (see previous section)\footnote{Moreover,
using an analogous approach, good agreement was also found with the corresponding data for jet multiplicites~\cite{Bolzoni:2012ii}.}.
Therefore, we believe that all these indicate toward success in describing the EMC ratio by using the same approach.

We note that the main difference between global fits and DAS approach is in the restriction of 
applicability of the latter by
low $x$ region only, while the advantage of the DAS approach lies in the analytic solution to DGLAP equations.

Thus, we are trying to apply the DAS approach to low $x$ region of EMC effect using a simple fact that the rise of parton
densities increases with increasing $Q^2$ values. This way, with scales of PDF evolutions
less than $Q^2$ (i.e. $\mu^2 \leq Q^2$) in nuclear cases, we can directly reproduce the shadowing effect which is observed
in the global fits. Since there are two components~(\ref{8.01}) for each parton density, we have two free parameters
$\mu_{\pm}$ to be fit in the analyses of experimental data for EMC effect at low $x$ values.

An application of the rescaling model at low $x$ can be incorporated at LO as follows:
\begin{eqnarray}
F^A_2(x,\mu^2) &=& e \, f^A_q(x,\mu^2),~~ F^N_2(x,\mu^2) = e \, f_q(x,\mu^2), 
\nonumber \\
  f^A_a(x,\mu^2) &=& 
%~=~ 
  f_a^{A,+}(x,\mu^2) + f_a^{A,-}(x,\mu^2),~~(a=q,g),~~
  %\nonumber \\
  f^{A,\pm}_a(x,\mu^2) =
  %  &=&
  f^{\pm}_a(x,\mu^2_{A,\pm}) \, ,
	\label{8.02A}
\end{eqnarray}
with a similar definition of $\mu^2_{A,\pm}$ as in the previous section (up to replacement
$v \to \pm$). The expressions for $f^{\pm}_a(x,\mu^2)$ are given in Eqs.~(\ref{8.01}) and~(\ref{8.02}).

Then, the corresponding values of $s^A_{\pm}$ are found to be
\begin{equation}
  s^A_{\pm} \equiv \ln \left(\frac{\ln\left(\mu^2_{A,\pm}/\Lambda^2\right)}{\ln\left(\mu^2_{0}/\Lambda^2\right)}\right)
  = s +\ln\Bigl(1+\delta^A_{\pm}\Bigr)\, ,
  %\approx s +\delta^A_{\pm},~~~ 
\label{sApm}
\end{equation}
because of the saturation at low $x$ values for all considered $Q^2$ values, which in our case should be related
with decreasing the arguments of ``$\pm$'' component. Therefore, the values of $\delta^A_{\pm}$ should be negative.

\section{Analysis of the low $x$ data for nucleus}

Note that it is usually convenient to study the following ratio
(see Fig.~1 in Ref.~\cite{Kulagin:2016fzf})
\begin{equation}
R^{AD}_{F2}(x,\mu^2) = \frac{F^A_2(x,\mu^2)}{F^D_2(x,\mu^2)}\,.
\label{AD}
\end{equation}

Using the fact that the nuclear effect in a deutron is very small (see Table~1 for the values
of $\delta^A_{v}$ and discussions in~\cite{Kulagin:2016fzf})
\footnote{The study of nuclear effects in a deutron can be found in~\cite{AKP},
  which also contains short reviews of preliminary investigations.},
we can suggest that
\begin{eqnarray}
  F^D_2(x,\mu^2) &=& e \, f_q(x,\mu^2),~~
  %\nonumber \\
  F^A_2(x,\mu^2) =
  %&=&
  e \,  \overline{f}^{A}_q(x,\mu^2),  \nonumber \\
\overline{f}^{A}_a(x,\mu^2) &=& 
%~=~ 
\overline{f}_a^{A,+}(x,\mu^2) + \overline{f}_a^{A,-}(x,\mu^2),~~(a=q,g),~~
%\nonumber \\
\overline{f}^{A,\pm}_a(x,\mu^2) =
%&=&
f^{\pm}_a(x,\mu^2_{AD,\pm}) \, ,
\label{AD1}
\end{eqnarray}
i.e.
\begin{eqnarray}
\overline{f}^{A,+}_g(x,\mu^2) &=& \biggl(A_g + \frac{4}{9} A_q \biggl)
I_0(\sigma^{AD}_{+}) \; e^{-\overline d_{+} s^{AD}_{+}}
+ O(\rho^{AD}_{+}), \nonumber \\
\overline{f}^{A,+}_q(x,\mu^2) &=& 
%&=& 
\frac{f}{9} \biggl(A_g + \frac{4}{9} A_q \biggl) \rho^{AD}_{+}
I_1(\sigma^{AD}_{+})  \; e^{-\overline d_{+} s^{AD}_{+}} + O(\rho^{AD}_{+}),
	\label{8.01AD} \\
%\nonumber \\
        \overline{f}^{A,-}_g(x,\mu^2) &=& -\frac{4}{9} A_q e^{- d_{-} s^{AD}_{-}} \,
        + \, O(x),~~
        %\nonumber \\
%	\label{8.00} \\
	\overline{f}^{A,-}_q(x,\mu^2) =  A_q e^{-d_{-}(1) s^{AD}_{-}} \, + \, O(x),
	\label{8.02AD}
\end{eqnarray}
where
\begin{eqnarray}
  &&\sigma^{AD}_{+}=\sigma(s\to s^{AD}_{+}),~~\rho^{AD}_{+}=\rho(s\to s^{AD}_{+}),~~ \nonumber \\
%  && \tilde{\sigma}^{AD}_{+}=\sigma^{AD}( s^{AD}_{+}\to - s^{AD}_{+}),~~  \tilde{\rho}^{AD}_{+}=\rho^{AD}( s^{AD}_{+}\to - s^{AD}_{+}),
%  \nonumber \\
  &&s^{AD}_{\pm} \equiv \ln \left(\frac{\ln\left(\mu^2_{AD,\pm}/\Lambda^2\right)}{\ln\left(\mu^2_{0}/\Lambda^2\right)}\right)
  = s +\ln\Bigl(1+\delta^{AD}_{\pm}\Bigr)
  %\approx s +\delta^{AD}_{\pm},~~ ~
  % \delta^{AD}_{\pm} = \delta^{A}_{\pm} - \delta^{D}_{\pm}
  \, .
%\label{AD2}
\end{eqnarray}

We obtain the values of $\delta^{AD}_{+}$ and $\delta^{AD}_{-}$
by fitting NMC experimenal data~\cite{Arneodo:1995cs}
for the EMC ratio at low $x$ in the case of different nuclei.
Since the experimental data for lithium and carbon are most precise
and contain the maximal number of points (16 points for each nucleus),
we preform combined fits of these data. 
Obtained results (with $\chi^{2}_{an}$=27 and  $\chi^{2}_{fr}$=43 for 32 points)
are presented in Table~3 and shown in Fig.~2.

\newpage
{\bf Table 3.}
%The difference ${\bf f}_u^v(2,Q^2)-{\bf f}_d^v(2,Q^2)$.
%\vspace{0.2cm}
\begin{center}
\begin{tabular}{|c|c|c|c|c|}
\hline
%        &               &    &  &                       &        &         \\
                & -$\delta^{AD}_{+,an}$ & -$\delta^{AD}_{-,an}$ & -$\delta^{AD}_{+,fr}$ & -$\delta^{AD}_{-,fr}$  \\
%&               &      &                           &    &          \\
\hline
   ${}^7$Li     &  0.061 $\pm$ 0.006 &  0.216 $\pm$ 0.065   & 0.073 $\pm$ 0.012  & 0.348 $\pm$ 0.067 \\
   ${}^{12}$C      &   0.105 $\pm$ 0.007   &  0.411 $\pm$ 0.042   & 0.139 $\pm$ 0.013  &   0.590 $\pm$ 0.041\\
\hline 
\end{tabular}
\end{center}

As can be seen in Fig.~2 there is large difference between the fits with ``frozen''
and analytic versions of the strong couling constant. This is in contrast with the analysis done in Section~1
and results done in the earlier papers~\cite{KoLiZo}.
It seems that this difference comes about because we include in the analysis the region of very low $Q^2$ values, 
where ``frozen'' and analytic strong couling constants are observed to be rather different (see also~\cite{Shirkov:2012ux}).

\begin{figure}[!hbt]
\centering
\vskip 0.5cm
\includegraphics[height=0.45\textheight,width=0.8\hsize]{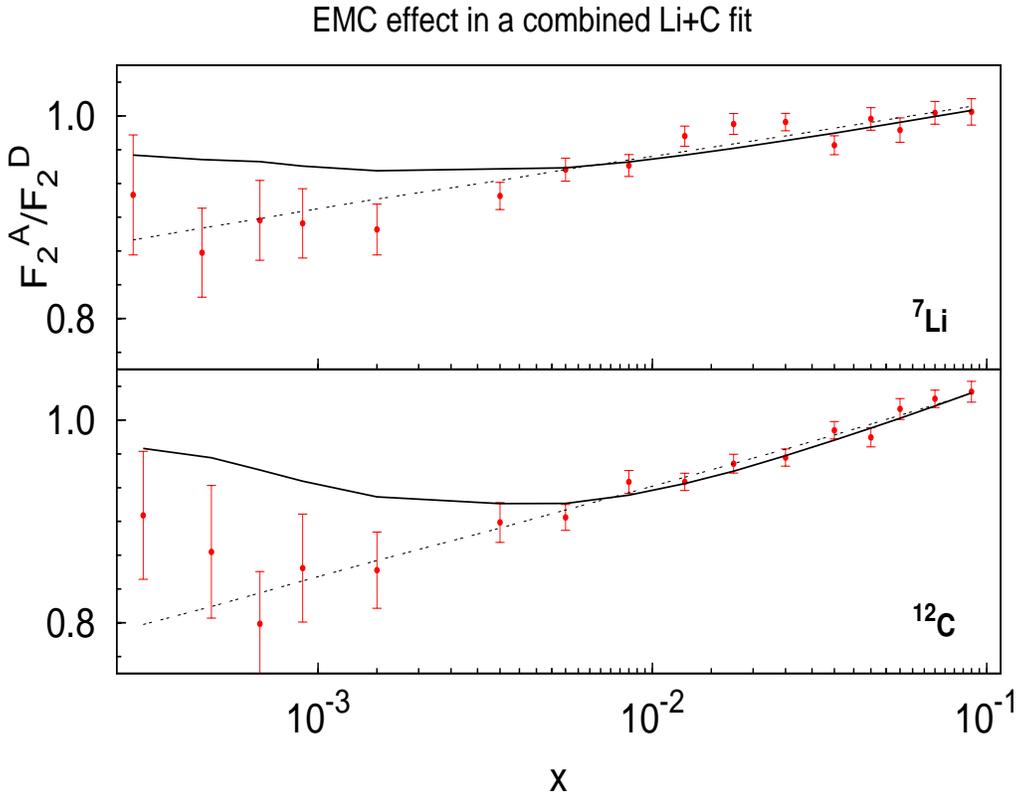}
\vskip -0.3cm
\caption{small $x$ dependence of $R^{AD}_{a}(x,\mu^2)$ for lithium and carbon.
  The combined experimental data from  NMC~\cite{Arneodo:1995cs}
%, FNAL and HERMES Collaborations~\cite{Aaron:2009aa}
are fitted by LO expressions
%for $Q^2\geq1$~GeV$^2$
implemented with the frozen (solid lines) and analytic (dashed lines)
modifications of the strong coupling constant.}
%\label{fig:F2}
\end{figure}

\vskip 0.5cm

\section{ $A$ dependence  at low $x$ }

Taking NMC experimental data~\cite{Arneodo:1995cs} along with E665 and HERMES Collaborations~\cite{Adams:1995is}
%(see Fig. 1 in  Ref. \cite{Kulagin:2016fzf})
for the EMC ratio at low $x$ in the case of different nuclei, we can find the $A$ dependence of $\delta^{AD}_{\pm}$,
which can be parameterized as follows
\be
- \delta^{AD}_{\pm} = c^{(1)}_{\pm} + c^{(2)}_{\pm} A^{1/3}.
\label{AD2}
\ee

As it was already mentioned in the previous section, usage of the analytic coupling constant leads 
to the fits with smaller $\chi^2$ values. For example, the values of $c^{(1)}_{\pm}$ and $c^{(2)}_{\pm}$ found
in the combined fit of the data (76 points) when the analytic coupling constant is used (with $\chi^2=89$)
look like
\bea
&& c^{(1)}_{+,an} = -0.055 \pm 0.015,~~ c^{(2)}_{+,an} = 0.068 \pm 0.006,~~ \nonumber \\
&& c^{(1)}_{-,an} = 0.071 \pm 0.101,~~ c^{(2)}_{-,an} = 0.120\pm 0.039 \, .
\eea \label{AD2.an}
% ~~ {\mbox with} ~~ \chi^2=27 \label{AD2.an} \\
%&& c^{(1)}_{+,fr} = 0.243 \pm 0.085,~~ c^{(2)}_{+,an} = -0.159\pm 0.040,~~ \nonumber \\
%&& c^{(1)}_{-,an} = 1.059 \pm 0.492,~~ c^{(2)}_{+,an} = -0.700\pm 0.223,
% ~~ {\mbox with} ~~ \chi^2=61 \label{AD2.fr}
%\eea
% with $\chi^2=89$.

Now, using the $A$ dependence (\ref{AD2}), $R^{AD}_{F2}(x,\mu^2)$ values for any nucleus $A$ can be predicted.
What is more, we can consider also the ratios $R^{AD}_{a}(x,\mu^2)$ of parton densities in a nucleus and deutron themselves,
\begin{equation}
R^{AD}_{a}(x,\mu^2) = \frac{\overline{f}^A_a(x,\mu^2)}{f_a(x,\mu^2)},~~ (a=q,g) \, ,
\label{ADa}
\end{equation}
with $\overline{f}^A_a(x,\mu^2)$ and ${f_a(x,\mu^2)}$ defined in Eqs.~(\ref{AD1})---(\ref{AD2}) and~(\ref{8.01})---(\ref{intro:1b}), respecively.

Indeed, at LO $R^{AD}_{q}(x,\mu^2)=R^{AD}_{F2}(x,\mu^2)$;  therefore, results for $R^{AD}_{q}(x,\mu^2)$ are already known.
Since all the parameters of PDFs found within the framework of the generalized DAS approach are now fixed
we can predict the ratio  $R^{AD}_{g}(x,\mu^2)$ of the gluon densities in a nucleus and nucleon given
in Eqs.~(\ref{8.01}), (\ref{8.02}), (\ref{8.01AD}) and (\ref{8.02AD}), which is currently under intensive studies 
(see a recent paper~\cite{Frankfurt:2016qca} and review~\cite{Armesto:2006ph} along with references and discussion therein).

\begin{figure}[t]
\centering
\vskip 0.5cm
\includegraphics[height=0.45\textheight,width=0.8\hsize]{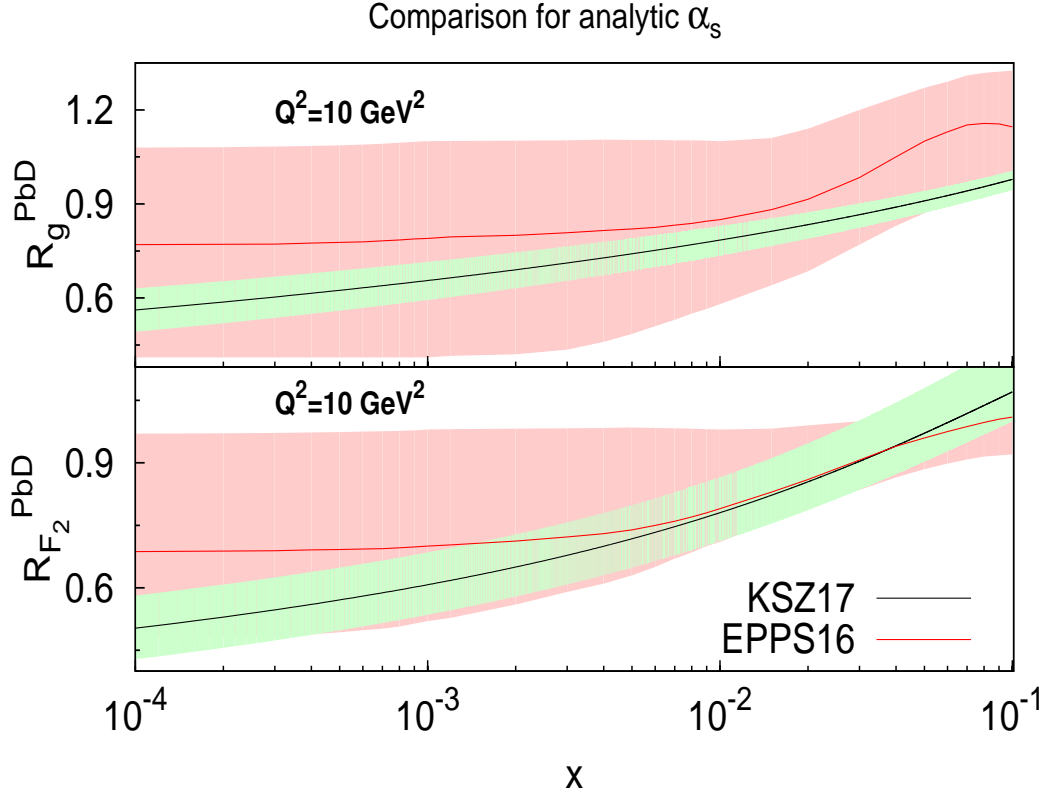}
\vskip -0.3cm
\caption{$x$ dependence of $R^{AD}_{F2}(x,\mu^2)$ and $R^{AD}_{g}(x,\mu^2)$ at $\mu^2$=10 GeV$^2$ for lead data. 
A green line with pink band (shows 90$\%$ uncertainties) is taken from the second paper of~\cite{Armesto:2006ph},
while a black one with light green band is obtained in the present paper.}
%\label{fig:F2}
\end{figure}

The results for $R^{AD}_{F2}(x,\mu^2)$ and $R^{AD}_{g}(x,\mu^2)$, depicted in Fig.~3, show some
difference between these ratios. It is also seen that the difference is similar to that
obtained in a recent EPPS16 analysis (see the first paper in~\cite{Eskola:2016oht})
\footnote{
Note that the result for $R^{AD}_{g}(x,\mu^2)$ along with its uncertainty is completely determined 
by both the rescaling model and the analytic form for parton densities at low $x$ values we've used.
Therefore, it is clear that the light green band for $R^{AD}_{g}(x,\mu^2)$ should become broader
due to a freedom in using various models.
Also note that a comparison between two uncertainty bands shown in Fig.~3 is in some sense misleading.
The pink band is much broader since the EPPS16 global analysis included a fit to all available data
across quite a wide range in $x$ as opposed to small $x$ consideration adopted in the present paper.
Nonetheless, we decided to quote it here just to give the reader an idea about the subject, at least qualitatively.}.
However, what for $R^{AD}_{F2}(x,\mu^2)$ and $R^{AD}_{g}(x,\mu^2)$ themselves (irrespective of other results),
we obtain a bit stronger effect at lowest $x$ values, which does in fact not contradict
the experimental data collected by the LHCb experiment (see recent review in~\cite{Winn:2017kwv}). Such a strong
effect is also well compatible with the leading order EPPS09 analysis (which can also be found in~\cite{Winn:2017kwv}). 
It will be interesting to delve into more in-depth studies of the 
ratio $R^{AD}_{g}(x,\mu^2)$, which is one of our aims in the future.

\section{SF $F_{2c}$ at low $x$}

Several years ago H1~\cite{Aaron:2009jy} and ZEUS~\cite{Chekanov:2009kj} Collaborations at HERA
have separately presented their new data on the charm structure function $F_{2c}$\footnote{Open charm production was
also observed in the COMPASS fixed target experiment~\cite{Adolph:2012ca}.} and more recently
they have combined these data on $F_{2c}(x,\mu^2)$~\cite{Abramowicz:1900rp}.
The SF $F_{2c}$ was found to be around 25\% of $F_{2}$, which is considerably larger than what was observed
by the European Muon Collaboration (EMC) at CERN \cite{Aubert:1982tt} at larger $x$ values, where it was only aroung 
1\% of  $F_{2}$.

Ensuing and very extensive theoretical analyses were carried out to establish that the $F_{2c}$ data can be described 
through the perturbative generation of charm in QCD \cite{Frixione:1994dv}.
In view of this, a PGF process in experiments with nucleon and nucleus targetsis one of the most 
effective and promising studies of gluon density (see a recent review~\cite{Chudakov:2016ytj}).

Following~\cite{Illarionov:2008be} the SF $F_{2c}$ at low $x$ can be represented
in the framework of the generalized DAS approach as follows
\be
F_{2c}(x,\mu^2) = e^2_c \, a_s(\mu_c) \, C_{2,g}(1, z_c(\mu^2)) f_g(x,\mu^2),~~~ z_c(\mu^2)=\frac{m^2_c(\mu^2)}{\mu^2},~~~ e_c=\frac{2}{3} \, ,
\label{c.1}
\ee
where $C_{2,g}(1, z_c(\mu^2)) $ is a first Mellin moment of the LO PGF coefficient function
%splitting function
$\tilde{C}_{2,g}(x, z_c(\mu^2))$. It can be obtained from the QED case~\cite{Baier:1966bf} by adjusting
the coupling constants
%and color factors and they read 
(see also the direct calculations in~\cite{Witten:1975bh,Kotikov:2001ct}). 
The Mellin moment  $C_{2,g}(1, z_c(\mu^2)) $ has a very compact form~\cite{Illarionov:2008be}:
\be
C_{2,g}(1, z) = \frac{2}{3} \, 
%a_s(\mu^2) 
\left[ 1- \frac{2(1-z)}{\sqrt{1+4z}} \, \ln 
\frac{\sqrt{1+4z}-1}{\sqrt{1+4z}+1} \right] \, .
\label{c.1a}
\ee
The gluon density $f_g(x,\mu^2)$ is determined in~(\ref{8.01}) and (\ref{8.02}).

The scale $\mu_c$ in~(\ref{c.1}) is actually not fixed because the results for $F_{2c}$ are at LO. There are two
widespread scales, $\mu_c^2=4m_c^2$ \cite{Chudakov:2016ytj,Gluck:1993dpa} and $\mu_c^2=4m_c^2+\mu^2$
\cite{Aaron:2009jy,Chekanov:2009kj,Abramowicz:1900rp,Illarionov:2008be}. We will use below both of them
(see Subsect.~7.1).

In the framework of the rescaling model the SF  $F^A_{2c}(x,\mu^2)$ for nucleus $A$ can be represented as follows
\be
F^A_{2c}(x,\mu^2) = e_c^2 \, \sum_{i=\pm} a_s(\mu_c(\mu^2_{A,i}))  \, C_{2,g}(1,z_c(\mu^2_{A,i})) f^{i}_g(x,\mu^2_{A,i})\,,
%~~~ a_c(\mu^2)=\frac{m^2_c(\mu^2)}{\mu^2)} \, ,
\label{c.2}
\ee
where the scale $\mu^2_{A,i}$ 
%and the coupling $a_s(\mu^2_{A,i})$ have 
looks like
\be
\mu^2_{A,\pm}= \Lambda^2 {\left(\frac{\mu^2}{\Lambda^2}\right)}^{1+\delta^A_{\pm}} =
\mu^2 {\left(\frac{\mu^2}{\Lambda^2}\right)}^{\delta^A_{\pm}}
%~~
 %a_s(\mu^2_{A,\pm}) = \frac{1}{\beta_0 \ln \left(\mu^2_{A,\pm}/\Lambda^2\right)} = 
 %\frac{a_s(\mu^2)}{1+\delta^A_{\pm}} 
 \, ,
\label{c.3}
\ee
as it follows from~(\ref{va.1}) with the replacement $v \to \pm$.

The results for the ratios $R^{A}_{F2}(x,\mu^2)$, $R^{A}_{g}(x,\mu^2)$ and 
\be
R^{A}_{c}(x,\mu^2)= \frac{F^A_{2c}(x,\mu^2)}{F_{2c}(x,\mu^2)}
%  \label{c.4}
\ee
should be rather similar. Moreover, they have  similar $x$-dependences, as it will be shown in the following subsection.
%, because the values of $\mu^2_{A,+}$ and  %$\mu^2_{A,-}$ are not much different from each other.

\subsection{Analysis of the low $x$ data}

To have as close a relation with analyses in Sect.~5 as possible, let us consider the ratio
\be
R^{AD}_{c}(x,\mu^2)= \frac{F^A_{2c}(x,\mu^2)}{F^D_{2c}(x,\mu^2)}\,.
  \label{c.4}
\ee
 
As in Sect.~5, we will use the following expressions for the SFs
 \bea
&& F^D_{2c}(x,\mu^2) = e^2_c \, a_s(\mu_c) \,C_{2,g}(1, a_c(\mu^2)) \, f_g(x,\mu^2), \nonumber \\
&& F^A_{2c}(x,\mu^2) =e_c^2 \,  \sum_{i=\pm} a_s(\mu_c(\mu^2_{AD,i}))  \, C_{2,g}(1, z_c(\mu^2_{AD,i})) \,
\overline{f}^{A,\pm}_g(x,\mu^2) \,,
% f^{i}_g(x,\mu^2_{AD,i}),
%~~~ a_c(\mu^2)=\frac{m^2_c(\mu^2)}{\mu^2)} \, ,
\label{c.6}
\eea
 where the gluon density $\overline{f}^{A,\pm}_a(x,\mu^2)= f^{\pm}_a(x,\mu^2_{AD,\pm})$ is defined
in~(\ref{8.01AD}) and (\ref{8.02AD}). The scale $\mu^2_{AD,\pm}$ 
 %and the coupling $a_s(\mu^2_{AD,i})$ 
 can be obtained from~(\ref {c.3}) with the replacement $\delta^A_{\pm} \to \delta^{AD}_{\pm}$,
by analogy with analyses in Sect.~5.
 
 \begin{figure}[t]
\centering
\vskip 0.5cm
\includegraphics[height=0.45\textheight,width=0.8\hsize]{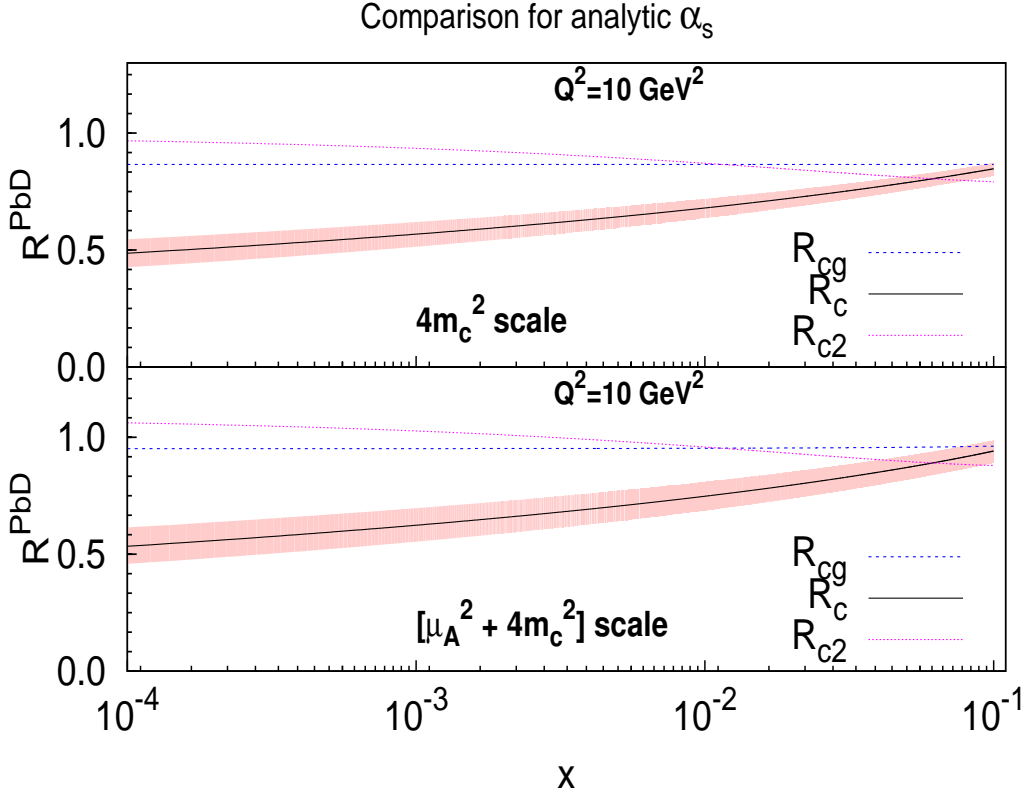}
\vskip -0.3cm
\caption{$x$ dependence of $R^{AD}_{c}(x,\mu^2)$, $R^{AD}_{cg}(x,\mu^2)$ and $R^{AD}_{c2}(x,\mu^2)$ at 
$\mu^2$=10 GeV$^2$ for lead data and  two choices of $\mu_c$ scale: $\mu_c^2=4m_c^2$ and $\mu_c^2=4m_c^2+\mu^2$
are shown by black, blue and pink lines, respectively. 
A band represents 90$\%$ level uncertainties in determining $R^{AD}_{c}(x,\mu^2)$ values.}
%\label{fig:F2}
\end{figure}
 
The results for the ratios $R^{AD}_{c}(x,\mu^2)$,
\be
 R^{AD}_{cg}(x,\mu^2)= \frac{R^{AD}_{c}(x,\mu^2)}{R^{AD}_{g}(x,\mu^2)}~~\mbox{ and }
 R^{AD}_{c2}(x,\mu^2)= \frac{R^{AD}_{c}(x,\mu^2)}{R^{AD}_{F2}(x,\mu^2)}
  \label{c.7}
\ee
 are presented in Fig.~4 for $\mu^2$ = 10 GeV$^2$. 
 Since the $\mu^2$-dependence of $m_c$ is not strong, we use fixed $m_c=1.27$ GeV~\cite{Breview}
 in our analysis.

As can be seen in Fig.~4, results look very much the same for both scales of $\mu_c$. 
What is more, a behavior of the ratio $R^{AD}_{c}(x,\mu^2)$ is a little bit weaker than
that of $R^{AD}_{F2}(x,\mu^2)$ and a bit stronger than that observed for $R^{AD}_{g}(x,\mu^2)$.
We hope that the $x$-dependence of the ratio $R^{AD}_{c}(x,\mu^2)$, along with that of $R^{AD}_{g}(x,\mu^2)$,
can be measured at a future Electron--Ion Collider (see~\cite{Chudakov:2016ytj} and discussion therein).

\section{Conclusion}

Using a recent progress in the application of double-logarithmic approximations~(see \cite{Q2evo,Kotikov:2012sm}
and~\cite{Bolzoni:2012ii}) to the studies of small $x$ behavior of the structure and fragmentation functions,
respectively, we applied the DAS approach~\cite{Munich,Q2evo} to examine an EMC $F_2$ structure function ratio 
between various nuclei and a deutron. Within a framework of the rescaling model~\cite{Close:1984zn,Close:1983tn}
good agreement between theoretical predictions and respective experimental data is achieved.

The theoretical formul\ae ~contain certain parameters, whose values were fit in
the analyses of experimental data. Once the fits are carried out we have predictions for 
the corresponding ratios of parton densities without free parameters. These results were used to
predict small $x$ behavior of the gluon density in nuclei, which is at present poorly known.

The ratios $R^{AD}_{a}(x,\mu^2)$ $(a=q,g)$ predicted in the present paper are compatible with those
given by various groups working in the area. From our point of view, it is quite valuable
that the application of the rescaling model~\cite{Close:1984zn,Close:1983tn} provided us with 
very simple forms for these ratios.
It should also be mentioned that without any free parameters we can predict the ratio $R^{AD}_{c}(x,\mu^2)$ 
of charm parts, $F^{A}_{2c}(x,\mu^2)$ and $F^{D}_{2c}(x,\mu^2)$, of the respective structure functions. 
This latter ratio
%$R^{AD}_{c}(x,\mu^2)$ 
has a simple form and it is very similar to the corresponding ratio of the complete structure functions
$F^{A}_{2}(x,\mu^2)$ and $F^{D}_{2}(x,\mu^2)$.

Following~\cite{Q2evo,Kotikov:2012sm} we plan to extend our analysis
to the NLO level of approximation, the accuracy that is currently a standard in nPDF studies.
Also, we are going to consider a rather broad range of the Bjorken variable $x$ by
using parametrizations of parton densities, which will be constructed by analogy 
with the one obtained earlier in the valence quark case (see~\cite{Illarionov:2010gy}).
The usage of such type of parametrizations will make it possible to carry out the present analysis of the data
accumulated within the range of intermediate $x$ values, which is presently under active considerations.\\

%\begin{acknowledgments}
Support by the National Natural Science Foundation of China (Grant
No. 11575254) is acknowledged. A.V.K. and B.G.S. thank Institute of Modern Physics 
for invitation. A.V.K. is also grateful to the CAS President's International Fellowship Initiative 
(Grant No.~2017VMA0040) for support.
The work of A.V.K. and B.G.S. was in part supported by the RFBR Foundation through the Grant No.~16-02-00790-a.
%\end{acknowledgments}

\end{document}